\documentclass[fleqn,usenatbib]{mnras}

\usepackage{newtxtext,newtxmath}

\usepackage[T1]{fontenc}
\usepackage{amsmath}
\usepackage{graphicx}
\usepackage{lastpage}
\usepackage[dvipsnames]{xcolor}   
\usepackage[super]{nth}
\usepackage{microtype}
\usepackage{booktabs, tabularx}
\usepackage{supertabular}
\usepackage{physics}
\usepackage[math]{cellspace}

\usepackage{hyperref}
\definecolor{royalblue}{rgb}{0.25, 0.41, 0.88}
\definecolor{darkviolet}{rgb}{0.58, 0.0, 0.83}

\title[Harrison-Zel'dovich Spectrum's comeback?]{Is the Harrison-Zel'dovich spectrum coming back? ACT preference for $n_s \sim 1$ and its discordance with Planck}

\author[W. Giar\`e, F. Renzi, O. Mena, E. Di Valentino, A. Melchiorri] {
	William Giar\`e,$^{1,2,5}$ \thanks{E-mail: william.giare@uniroma1.it}
	Fabrizio Renzi,$^{3}$ \thanks{E-mail: renzi@lorentz.leidenuniv.nl}
	Olga Mena,$^{4}$ \thanks{E-mail: omena@ific.uv.es}
	Eleonora Di Valentino,$^{5}$ \thanks{E-mail: e.divalentino@sheffield.ac.uk}
	Alessandro Melchiorri,$^{2,6}$ \thanks{E-mail: alessandro.melchiorri@roma1.infn.it}
	\\
$^{1}$ Galileo Galileo Institute for theoretical physics, Centro Nazionale INFN di studi avanzati, Largo Enrico Fermi 2,  I-50125, Firenze, Italy\\
$^{2}$ INFN Sezione di Roma, P.le A. Moro 2, I-00185, Roma, Italy\\
$^{3}$ Lorentz Institute for Theoretical Physics, Leiden University, PO Box 9506, Leiden 2300 RA, The Netherlands\\
$^{4}$ IFIC, Universidad de Valencia-CSIC, 46071, Valencia, Spain\\
$^{5}$ School of Mathematics and Statistics, University of Sheffield, Hounsfield Road, Sheffield S3 7RH, United Kingdom\\
$^{6}$ Physics Department, Universit\`a di Roma ``La Sapienza'', Ple Aldo Moro 2, 00185, Rome, Italy}
    \pubyear{2021}
    
\begin{document}
\label{firstpage}
\pagerange{\pageref{firstpage}--\pageref{lastpage}}
\maketitle

\begin{abstract} 
The Data Release 4 of the Atacama Cosmology Telescope (ACT) shows an agreement with an Harrison-Zel'dovich primordial spectrum ($n_s=1.009 \pm 0.015$), introducing a tension with a significance of $99.3\%$~CL with the results from the \emph{Planck} satellite. The discrepancy on the value of the scalar spectral index is neither alleviated with the addition of large scale structure information nor with the low multipole polarization data. We discuss possible avenues to alleviate the tension relying on either neglecting polarization measurements from ACT or in extending 
different sectors of the theory.
\end{abstract}
    	
\begin{keywords}
cosmology: observations -- cosmology: theory -- cosmological parameters -- inflation
\end{keywords}
    	
\section{introduction}

Inflation provides the most successful cosmological scenario able to generate the initial conditions of our Universe and simultaneously solving the standard cosmological problems. However, and despite this remarkable success, the inflationary paradigm is still lacking firm observational confirmation. 

A "smoking-gun" evidence for inflation would be the detection of primordial B-modes in the Cosmic Microwave Background (CMB) power spectrum produced by primordial gravitational waves. In the most typical inflationary models, the amplitude of tensor perturbations is expected to be proportional to the quantity $|n_{\rm s}-1|^2$ with $n_{\rm s}$ the scalar spectral index of the primordial scalar spectrum:  the larger the departure of $n_{\rm s}$ from unity, the more likely tensor modes would be within observational reach. Therefore determining how much the former index deviates from one dictates the theoretical, phenomenological, also experimental perspectives of the field. For instance, a cosmological model with $n_{\rm s}=1$ -- that will corresponds to the phenomenological model proposed by Harrison, Zel’dovich, and Peebles~\citep{Harrison:1969fb,Zeldovich:1972zz,Peebles:1970ag} -- will imply a major theoretical breakthrough, as it would imply that the origin of cosmic perturbations may lie in some unknown fundamental theory different from the standard inflationary picture or in extensions of the latter~\citep{Barrow:1990vx,Barrow:1990td,Barrow:1993zq,Vallinotto:2003vf,Starobinsky:2005ab,Barrow:2006dh,delCampo:2007iw,Takahashi:2021bti,Ye:2022efx,Lin:2022gbl}.

In this regard, the latest observations of the CMB temperature and polarization anisotropies, echoes of the Big Bang, provided by the \emph{Planck} satellite have reached sub-percent accuracy on the extraction of the majority of the cosmological parameters~\citep{Planck:2018vyg,Planck:2018nkj}, resulting in a $\sim 8\sigma$ evidence for $n_{\rm s}\ne 1$ and establishing inflation as the most accredited theory of the early Universe. However, this is both a blessing and a curse, since, as high precision parameter extraction becomes a reality, the possible discrepancies among different data sets may grow in significance. 

Currently, there are several anomalies that can not be fully understood in the minimal cosmological constant plus cold dark matter ($\Lambda$CDM) scenario~\citep{Abdalla:2022yfr,Perivolaropoulos:2021jda}. The most significant $5\sigma$ disagreements is related to the value of the Hubble constant $H_0$ extracted from local distances and redshifts in the nearby Universe and that inferred from CMB observations~\citep{Riess:2021jrx,Verde:2019ivm,DiValentino:2020zio,DiValentino:2021izs}. Other less significant disagreements concern the parameter $S_8$, whose values differ for CMB and weak lensing estimates~\citep{DiValentino:2020vvd,Heymans:2020gsg,DES:2021wwk}, and the so-called lensing anomaly~\citep{Planck:2018vyg,Motloch:2018pjy}, related to the fact that the \emph{Planck} CMB data show a \emph{preference} for additional lensing.  Interestingly, while inflation predicts a perfectly flat Universe,  this excess of lensing in the damping tail produces an indication for a closed Universe at level of 3.4 standard deviations~\citep{Planck:2018vyg,Handley:2019tkm,DiValentino:2019qzk,DiValentino:2020hov,Semenaite:2022unt} that, if confirmed, would be very hard to explain in the simplest models of inflation. On the other hand, \emph{Planck}-independent small scale CMB observations provided by the Atacama Cosmology Telescope (ACT) and South Pole Telescope (SPT) fully support the inflationary prediction of a flat Universe, suggesting that the \emph{Planck} curvature anomaly may be due to a statistical fluctuation or an undetected systematic. Consequently, also the lensing amplitude aligns with the expected values in the $\Lambda$CDM model. Nonetheless, ACT and SPT data in turn show other mild yet relevant deviations from the $\Lambda$CDM scenario~\citep{DiValentino:2022rdg, DiValentino:2022oon, Calderon:2023obf}, including a $\sim 2.7\sigma$ discrepancy in the scalar spectral index $n_{\rm s}$ between \emph{Planck} ($n_{\rm s}=0.9649\pm 0.0044$) and ACT ($n_{\rm s}=1.008\pm 0.015$) that represents a new potential challenge for inflationary cosmology. As with the other tensions mentioned above, this $n_{\rm s}$ discrepancy could result from a statistical fluctuation, to a (yet unknown) systematic effect in the ACT or \emph{Planck} data, or a departure from the theoretical $\Lambda$CDM framework by assuming canonical inflation as the dominant mechanism for producing the perturbations in the early Universe. In this regard, recent analyses~\citep{DiValentino:2018zjj,Ye:2022efx,Jiang:2022uyg,Jiang:2022qlj} suggest a potential prominent role of $n_s$ in solving the aforementioned cosmological tensions, motivating the need of a systematic investigation on the nature of this discrepancy. In this paper we therefore scrutinize the emergent tension on canonical inflationary scenarios, exploring different cosmological observations at distinct epochs in the cosmic evolution to evaluate their robustness. The work is structured as follows: \autoref{sec.2} outlines our methodology and the data-sets utilized throughout this study. In \autoref{sec.3}, we conduct a comprehensive re-analysis of the ACT and \emph{Planck} data, extending the discussion presented in \cite{ACT:2020gnv} by incorporating additional CMB observations and updated large-scale structure data. In \autoref{sec.4} we explore the possible reasons behind the emerging tension by evaluating many different theoretical scenarios beyond the standard cosmological model and highlighting intriguing avenues. Finally, \autoref{sec.5} concludes with our findings.

\section{Methods}
\label{sec.2}
We employ the Monte Carlo Markov Chain (MCMC) method to analyze the posterior distributions of our parameter space, utilizing the publicly available \textsc{COBAYA} software~\citep{Torrado:2020xyz}. The MCMC sampler used in the analysis has been adapted from \textsc{CosmoMC} \citep{Lewis:2002ah} and incorporates the "fast dragging" procedure detailed in~\cite{Neal:2005}. The theoretical models are calculated using the latest version of the cosmological Boltzmann integrator code \textsc{CAMB}~\citep{Lewis:1999bs,Howlett:2012mh}. Our prior distributions for the parameters are uniform, with the exception of the optical depth $\tau$, which is selected based on CMB datasets as detailed below.

Our main datasets consist of the observations of the Cosmic Microwave Background provided by the \emph{Planck} satellite and the Atacama cosmology Telescope. In particular we use

\begin{itemize}

\item The full \emph{Planck} 2018 temperature and polarization likelihood~\citep{Planck:2019nip,Planck:2018vyg,Planck:2018nkj}, including multipoles $30\lesssim \ell \lesssim 2500$ for the TT, TE, EE spectra and low multipole data $2\le \ell \le 30$ for the EE spectrum. We refer to this dataset as "Planck".
\\
\item The \emph{Planck} 2018 temperature and polarization likelihood~\citep{Planck:2019nip,Planck:2018vyg,Planck:2018nkj}, including only low multipoles $30\lesssim \ell \lesssim 650$ for the TT, TE, EE spectra and $2\le \ell \le 30$ for the EE spectrum. We refer to this dataset as "Planck $(2\le \ell \le 650 )$".
\\
\item The \emph{Planck} 2018 temperature and polarization likelihood~\citep{Planck:2019nip,Planck:2018vyg,Planck:2018nkj}, including only high multipoles $\ell > 650$ for the TT, TE, EE spectra and $2\le \ell \le 30$ for the EE spectrum. We refer to this dataset as "Planck $(\ell > 650)$".
\\
\item The \emph{Atacama Cosmology Telescope} TT TE EE DR4 likelihood~\citep{ACT:2020frw}, both assuming a conservative Gaussian prior on $\tau=0.065\pm0.015$ and assuming a \emph{Planck}-based prior on $\tau=0.0544\pm0.070$. We refer to these two datasets as "ACT" and "ACT ($\tau=0.0544\pm0.070$)", respectively.
\\
\item The \emph{Atacama Cosmology Telescope} TT TE EE DR4 likelihood~\citep{ACT:2020frw}, in combination with \emph{Planck} low multipole polarization measurements $2\le \ell \le 30$ for the EE spectrum. We refer to this dataset as "ACT+Planck lowE".
\\
\item  The \emph{Atacama Cosmology Telescope} TT DR4 likelihood~\citep{ACT:2020frw}, in combination with the South Pole Telescope TE EE polarization measurement and a gaussian prior $\tau=0.065\pm0.015$. We refer to this dataset as "ACT+SPT".
\end{itemize}

In addition to CMB observations, we utilize a variety of large scale structure data to complement our analysis:

\begin{itemize}
\item The Baryon Acoustic Oscillations (BAO) and Redshift Space Distortions (RSD) measurements from BOSS DR12~\citep{BOSS:2012dmf}. We refer to this dataset as "BAO (DR12)".
\\
\item The Baryon Acoustic Oscillations (BAO) and Redshift Space Distortions (RSD) measurements from eBOSS~\citep{Dawson:2015wdb}. We refer to this dataset as "BAO (DR16)".
\\
\item The shear-shear, galaxy-galaxy, and galaxy-shear correlation functions from the first year of the \emph{Dark Energy Survey}~\citep{DES:2017myr}. We refer to this dataset as "DES".
\end{itemize}

Regarding the theoretical model, in \autoref{sec.3} we mainly focus on the standard $\Lambda$CDM model with its  six canonical parameters. However, to assess the reliability of this emergent discrepancy in the value of $n_s$ and understand its relationship with other anomalous parameters, in \autoref{sec.4} we extend the baseline cosmology by modifying the reionization epoch, the neutrino sector (parametrized by the total neutrino mass $\sum m_{\nu}$ and the effective number of relativistic neutrinos at recombination $N_{\rm eff}$), the lensing amplitude ($A_{\rm lens}$), the curvature ($\Omega_k$), the dark energy equation of state ($w$), and the inflationary sector (allowing a running of the scalar tilt $\alpha_s$).

\section{Analysis}
\label{sec.3}
\begin{table}%
\centering
\renewcommand{\arraystretch}{1.5}
\begin{tabular}{l @{\hspace{0.4 cm}} c }
\toprule
\textbf{Dataset} & \textbf{Scalar Spectral Index} (\boldmath{$n_s$})  \\
        & \boldmath{$\Lambda$}\textbf{CDM}     \\
\hline\hline
		ACT                                 &  $1.009\pm 0.015$          \\
            ACT ($\tau=0.0544 \pm 0.0070$) &                    $1.007\pm 0.015$ \\
            ACT + Planck low E                        & $1.001\pm 0.011$       \\
		ACT+BAO (DR12)                              &  $ 1.006 \pm 0.013 $          \\
		ACT+BAO (DR16)                             &  $1.006\pm 0.014$          \\
		ACT+DES                             &  $1.007\pm 0.013$          \\
            ACT+SPT+BAO (DR16)                    &  $0.997\pm 0.013$        \\
		ACT+SPT+BAO (DR12)                          &  $0.996\pm 0.012$          \\
		\hline
		Planck                                &  $0.9649\pm 0.0044$        \\
		Planck+BAO (DR12)                           &  $0.9668\pm 0.0038$        \\
		Planck+BAO (DR16)                           &  $0.9677\pm 0.0037$        \\
            Planck+DES                              &  $0.9696\pm 0.0040$
            \\
		Planck ($2\le \ell \le 650$)          &  $0.9655\pm 0.0043$        \\
		Planck ($\ell > 650$)                 &  $0.9634\pm 0.0085$        \\

\bottomrule
    \end{tabular}
    \caption{\small The marginalized 1$\sigma$ bounds for the scalar spectral index for various data combination obtained assuming a standard cosmological model.}
\label{tab.Results.LCDM}
\end{table}

\begin{figure}
\includegraphics[width=\columnwidth]{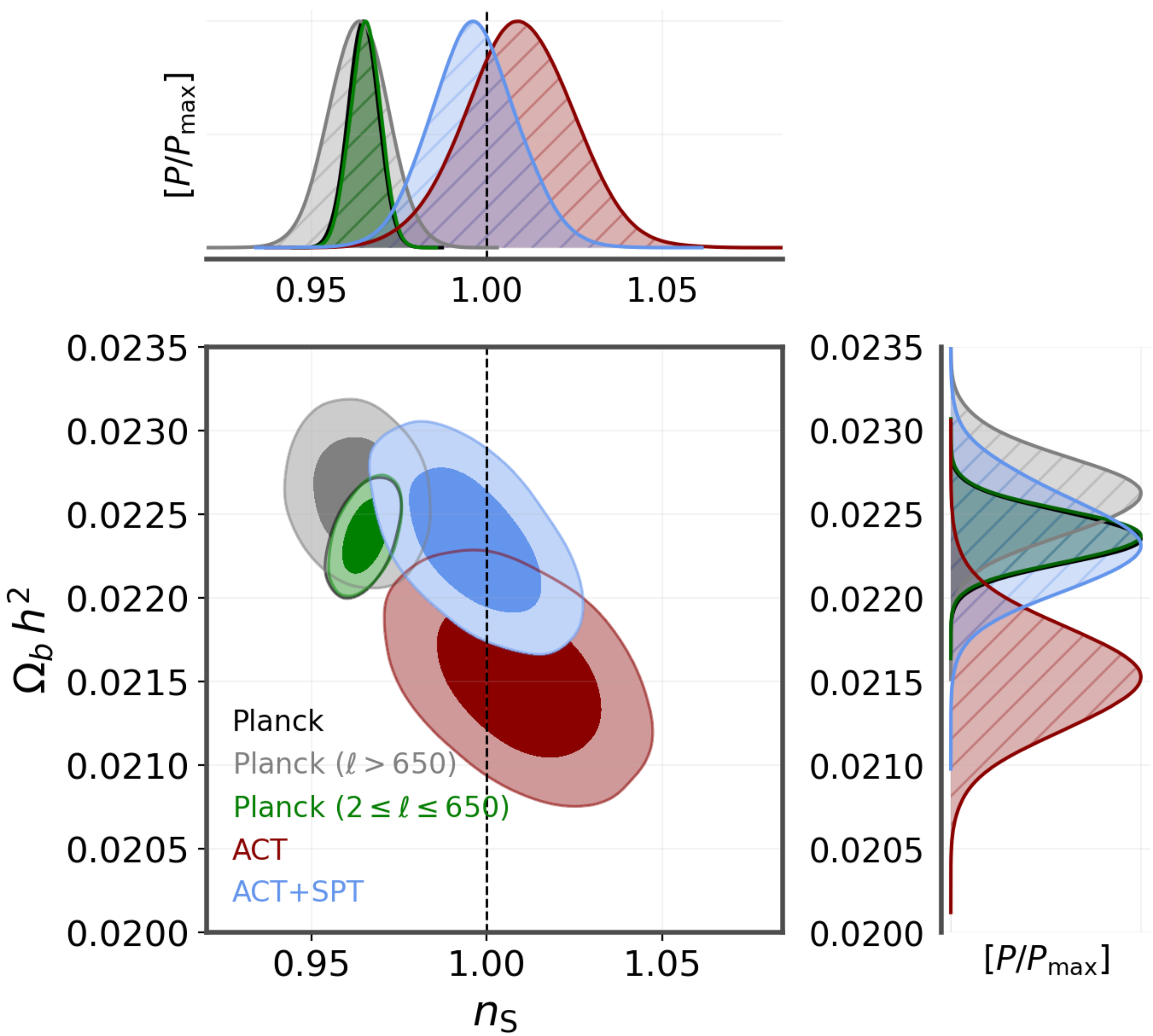}
\caption{\small One-dimensional posterior distributions and two-dimensional joint marginalized contours inferred by combining the Atacama Cosmology Telescope measurements of the CMB temperature anisotropies and the South Pole Telescope measurements of the CMB polarization anisotropies and splitting the \emph{Planck} likelihood in low ($2\le\ell\le650$) and high ($\ell>650$) multipoles.}
\label{fig:fig1}
\end{figure}

All our results for the $\Lambda$CDM model are summarized in~\autoref{tab.Results.LCDM}. As already mentioned, considering CMB data alone, the measurements of $n_{\rm s}$ from \emph{Planck} ($n_{\rm s}=0.9649\pm 0.0044$) 
and from ACT ($n_{\rm s}=1.008\pm 0.015$) differ by $\sim 2.7 \sigma$. This can be clearly observed in \autoref{fig:fig1} (see also Figure 14 of Ref.~\citep{ACT:2020gnv} for a comparison), where in the two-dimensional plane it can be definitely noted that the direction of the $\Omega_b h^2$-$n_{\rm s}$ degeneracy is opposite for ACT and \emph{Planck}, and the disagreement here is significantly exceeding $\sim 3 \sigma$. 
In the absence of low-$\ell$ CMB data, as in the case of ACT measurements, there is a strong degeneracy between the baryon energy density $\Omega_b h^2$ and the scalar spectral index $n_{\rm s}$: a lower value of the former (increasing the damping of the low $\ell$ acoustic peaks) can be always mimicked by a larger value of the latter, tilting the spectrum in the opposite direction. Our analysis therefore confirms that ACT measurements of the small scale CMB spectra favor a cosmology with a lower value of $\Omega_b h^2$ and a higher spectral index. For instance when fitting the ACT data, fixing the value of the spectral index to the \emph{Planck} measured value $n_{\rm s}=0.9649$ would give a larger $\chi^2=286.6$ than fixing $n_{\rm s}=1$ ($\chi^2=279.0$). As a result, ACT prefers a lower amplitude of the first acoustic peak in the TT power spectrum than both the \emph{Wilkinson Microwave Anisotropy Probe} (WMAP)~\citep{WMAP:2012nax} and \emph{Planck} CMB observations. Indeed, in Ref.~\citep{ACT:2020gnv} the mismatch in the values of $n_{\rm s}$ was interpreted as a consequence of the lack of information concerning the first acoustic peak of the temperature power spectrum. To verify this origin of the discrepancy in the CMB values of $n_{\rm s}$, we have performed two separate analyses of the \emph{Planck} observations, splitting the likelihood into low  ($2 \le \ell \le 650$) and high ($\ell > 650$) multipoles. We find that the discrepancy still persists at the level of $3\sigma$ ($2\sigma$) for low (high) multiple temperature data.
Our results therefore cast doubts on the claim that the mismatch in $n_s$ between ACT and \emph{Planck} is due to the lack of information on the first acoustic peak in ACT data. In fact, \emph{Planck} data still prefers a value of the scalar spectral index smaller than unity at $\sim 4.3 \sigma$ when the information on the first acoustic peak is removed, see \autoref{fig:fig1}. In addition, by focusing only on the high-$\ell$ region of the \emph{Planck} spectra, the disagreement with ACT is actually reduced at the level of $2\sigma$, but this is due to a loss of constraining power rather than a true shift of the mean value of $n_{s}$, see also \autoref{fig:fig1}. Therefore this discrepancy, although minor, should be seriously taken into account, as one would expect a reasonable agreement between two experiments measuring an overlapping range of multipoles. Conversely the low-$\ell$ end of the \emph{Planck} data is in strong disagreement with ACT: even with larger error bars, the low-$\ell$ data exhibits a tension higher than when the full \emph{Planck} multipole range is used.

One possible logical first step is to identify which of the data sets could be responsible for the $n_{\rm s}$ discrepancy, and discard it in the cosmological parameter inference analyses. We have made a number of tests along this line. In particular, in \cite{ACT:2020gnv} it was argued that an overall TE calibration could eventually explain the mismatch in $n_s$. We have therefore neglected any information arising from ACT polarization measurements (TE EE) and combined ACT temperature anisotropies (TT) with SPT polarization data (TE EE). In this case the disagreement with \emph{Planck} is reduced below $2\sigma$ (see also \autoref{fig:fig1}), but with the ACT and SPT data still preferring a value of $n_s$ around unity. However the tension is only mitigated by the larger error bars and once BAO are combined with the ACT and SPT data the disagreement with \emph{Planck} in fact grows again to the statistical level of $\sim 2.4 \sigma$ ($n_{\rm s}=0.996\pm 0.012$). Nonetheless, the combination of the ACT temperature and the SPT polarization produces a significant shift in the plane ($n_s$, $\Omega_b\,h^2$) resulting in a value of the baryon energy density $\Omega_b h^2=0.02237\pm 0.00030$ that is now in perfect agreement with the \emph{Planck} result. This result is significant as it shows that the degeneracy between the two parameters only partially contributes to the potential tension: restoring the agreement for $\Omega_b\,h^2$ may not be enough to reconcile the $n_s$ discrepancy.

\begin{figure}
\includegraphics[width=\columnwidth]{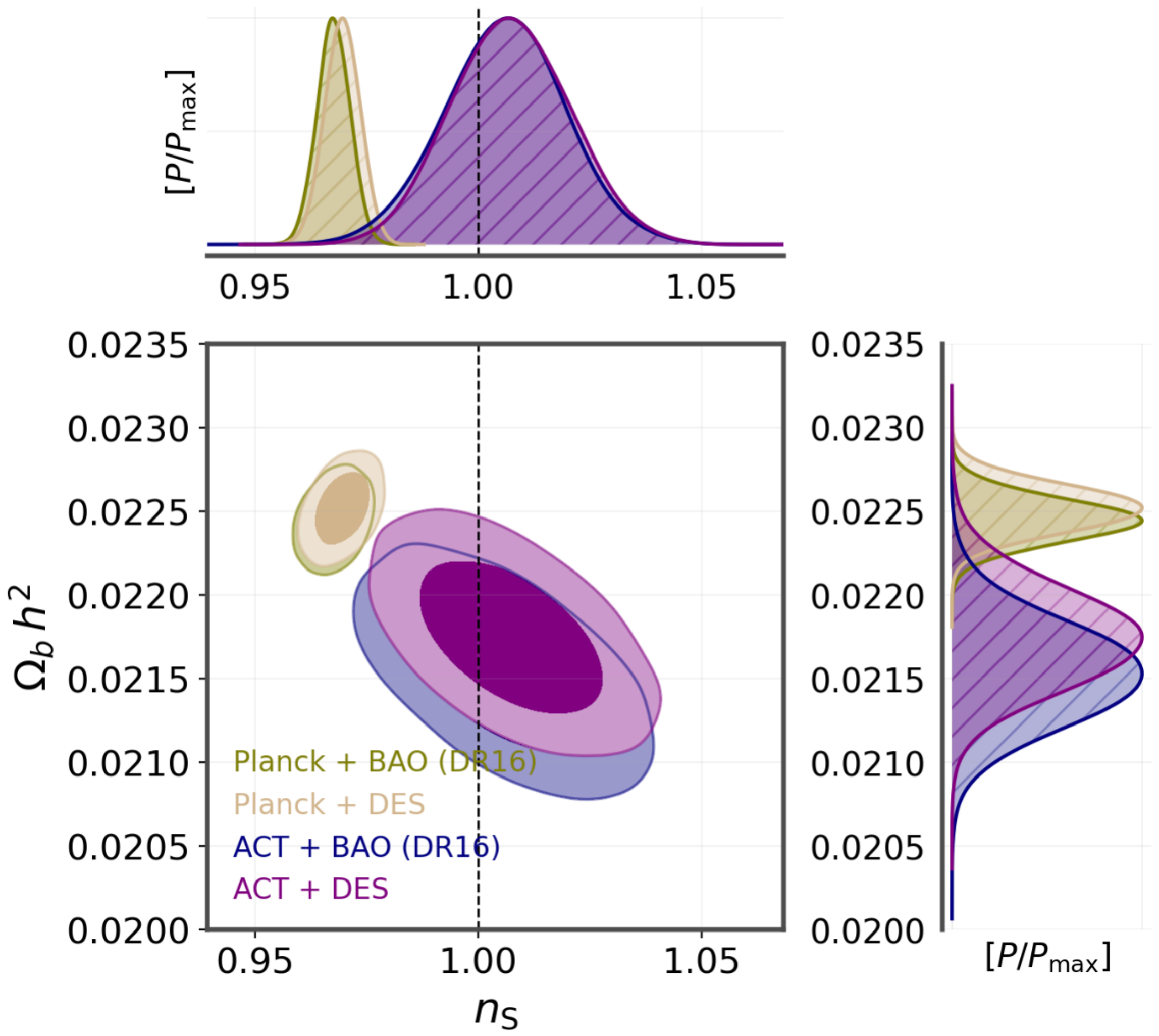}
\caption{\small One-dimensional posterior distributions and two-dimensional joint marginalized contours inferred by the Atacama Cosmology Telescope and \emph{Planck} satellite measurements, in conjunction with large scale structure data from BAO and DES observations.}
\label{fig:fig2}
\end{figure}

The next logical step is to investigate the effect of complementary (i.e.~non-CMB) data: the addition of BAO measurements normally alleviates tensions and restores the parameter values to those corresponding to $\Lambda$CDM. One well-known example is that of the curvature $\Omega_k$: the addition of BAO measurements to  \emph{Planck} observations is indeed very consistent with a flat cosmology ($\Omega_k= 0.0007 \pm 0.0019$ at $68\%$~CL). Unfortunately, this is not the case here: an inspection into  \autoref{fig:fig2} clearly shows that neither the addition of BAO DR16 nor that of BAO DR12, can alleviate the tension in the measured value of $n_{\rm s}$. As also noted in \cite{ACT:2020gnv}, combining ACT and BAO does not result in a noticeable shift in the parameter-space, but it does lead to tighter constraints on the parameters. As a result, the inclusion of both BAO measurements and a prior on the reionization optical depth leads to a tension even more significant, as the mean value of $n_{\rm s}$ remains unchanged but their error bars are reduced. Building upon the tests conducted in ~\cite{ACT:2020gnv}, we have expanded our analysis by incorporating the combination of ACT with DES galaxy clustering and cosmic shear observations. Our results summarized in \autoref{fig:fig2} show a similar tension, with a significance level of $3.1\sigma$. This highlights the persistence of the discrepancy even with the addition of different large scale structure datasets and further underscores the importance of understanding its underlying causes.

\section{Discussion}
\label{sec.4}

From all our analyses and consistency checks detailed above the lesson we learned is that there is a clear discrepancy between the extraction of the scalar spectral index of primordial perturbations from \emph{Planck} and ACT data. Since future strategies in B modes searches in the CMB polarization pattern depend crucially on the amplitude of these fluctuations, it is mandatory to understand where this anomaly comes from and what we can conclude about the precise value of $n_{\rm s}$ from present cosmological observations~\citep{Jiang:2022qlj}. In this section, we conduct further tests and explore possible ways to relieve the tension by examining theoretical extensions to the standard $\Lambda$CDM framework. Our aim is to identify potential modifications to the current model that may reconcile the observed discrepancies and provide a more comprehensive understanding of its origin.

\subsection{Extending the Reionization epoch}
\begin{figure*}
\includegraphics[width=\textwidth]{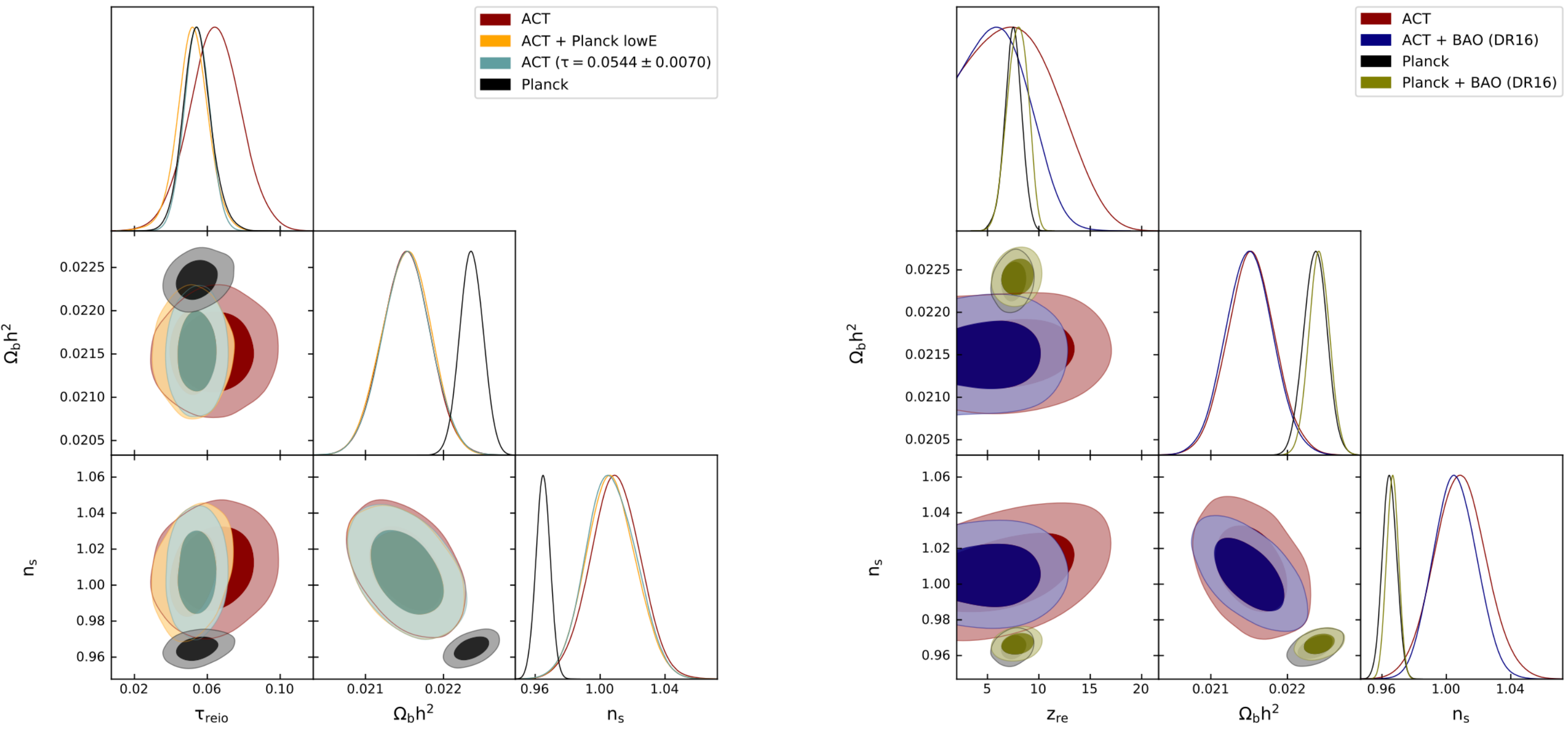}
\caption{\small One-dimensional posterior distributions and two-dimensional joint marginalized contours inferred by the Atacama Cosmology Telescope and \emph{Planck} satellite measurements by adopting a canonical parameterization for the reionization epoch (left panel) and considering the extended $(z\, , \,\Delta z)$ parameterization (right panel), respectively.}
\label{fig:fig3}
\end{figure*}

In order to understand the nature of this anomaly and recover the measured value of the scalar spectral index $n_{\rm s}$ by \emph{Planck} consistent with the predictions from most of the canonical inflationary scenarios, one could follow many distinct avenues. We start focusing on studying the impact of a more general reionization scenario on the tension on $n_{\rm s}$. The left Panel of \autoref{fig:fig3} illustrates that there is also a degeneracy between the reionization optical depth $\tau_{\rm reio}$ and $n_{\rm s}$. The reionization optical depth is defined as:
\begin{equation}
\tau_{\rm reio}(z) = \int_z^{\infty} dz' \frac{c ~dt'}{dz'}  (n_{\rm e}(z')- n_{\rm e, 0}(z'))\sigma_{\rm T}\,~,
\label{eq:reio}
\end{equation}
where $n_{\rm e}(z)=n_{\rm H}(0)(1+z)^3x_{\rm e}(z)$ and $n_{\rm e,0}(z)=n_{\rm H}(0)(1+z)^3x_{\rm e, 0}(z)$, being $n_{\rm H}(0)$ the number density of hydrogen at present, $x_{\rm e}(z)$ the free electron fraction and $x_{e,0}(z)$ the free electron fraction leftover from the recombination epoch. It is well-known that the statement that $n_{\rm s} = 1$ is observationally excluded no longer applies if one treats reionization in a general manner~\citep{Pandolfi:2010dz}. One could therefore consider to add an additional prior on the reionization optical depth to the ACT constraints on the cosmological parameters.  If a prior on $\tau_{\rm reio}=0.065 \pm 0.015$~\citep{ACT:2020gnv} ($\tau_{\rm reio}=0.0544\pm 0.0070$~\citep{Planck:2018vyg}) is applied, we have $n_\mathrm{s} = 1.009\pm 0.015$ ($1.007\pm 0.015$), barely changing the $\sim 3\sigma$ discrepancy with the \emph{Planck} results. The same conclusion is reached if ACT data is directly combined with the low-$\ell$ polarization \emph{Planck} (lowE) data, see also the left panel of~\autoref{fig:fig3}.

\begin{table}%
\centering
\renewcommand{\arraystretch}{1.5}
\begin{tabular}{l @{\hspace{0.4 cm}} c  @{\hspace{0.4 cm}}  c}
\toprule
\textbf{Model} & \textbf{Planck} (\boldmath{$n_s$})  &  \textbf{ACT} (\boldmath{$n_s$})\\
\hline\hline
$\Lambda\text{CDM} + z +\Delta z$ & $0.9647\pm 0.0044$ & $1.009\pm 0.016$ \\
$\Lambda\text{CDM}+A_{\text{lens}}$ & $0.9708\pm 0.0048$ & $1.008\pm 0.017$\\
$\Lambda\text{CDM}+N_{\text{eff}}$ & $0.9597\pm 0.0085$ & $0.960\pm 0.035$ \\
$\Lambda\text{CDM}+\Omega_{k}$ & $0.9706\pm 0.0048$ & $1.007\pm 0.016$ \\
$w\text{CDM}$ &$0.9654\pm 0.0043$ & $1.007\pm 0.016$\\
$\Lambda\text{CDM}+\sum m_{\nu}$ & $0.9646\pm 0.0044$ & $0.990^{+0.022}_{-0.019}$ \\
$\Lambda\text{CDM}+\alpha_s$ & $0.9635\pm 0.0046
$ & $0.980\pm 0.020$\\
$w\text{CDM}+\Omega_{k}$ &$0.9708\pm 0.0047$ & $1.007\pm 0.017$ \\
$\Lambda\text{CDM}+\Omega_{k}+\sum m_{\nu}$ & $0.9688\pm 0.0050$ & $0.987\pm 0.019$ \\
$w\text{CDM}+\Omega_{k}+\sum m_{\nu}$ &$0.9691\pm 0.0051$ & $0.986\pm 0.019$ \\
$w\text{CDM}+\Omega_{k}+\sum m_{\nu}+N_{\text{eff}}$ & $0.9686\pm 0.0095$ & $0.928\pm 0.045$\\
$w\text{CDM}+\Omega_{k}+\sum m_{\nu}+\alpha_s$ & $0.9689\pm 0.0054$ & $0.920\pm 0.031$\\
$w\text{CDM}+\Omega_{k}+N_{\text{eff}}+\alpha_s$ & $0.967\pm 0.012$ & $0.934\pm 0.050$\\
$w\text{CDM}+\sum m_{\nu}+N_{\text{eff}}+\alpha_s$ & $0.951\pm 0.011$ & $0.928\pm 0.033$\\
$w\text{CDM}+\Omega_{k}+\sum m_{\nu}+N_{\text{eff}}+\alpha_s$ & $0.968\pm 0.012$ & $0.943\pm 0.043$\\
\bottomrule
    \end{tabular}
    \caption{\small The marginalized 1$\sigma$ bounds for the scalar spectral index for various extensions of the cosmological model  as inferred by \emph{Planck} and ACT.}
\label{tab.Results.Models}
\end{table}

A different avenue is to relax the reionization scenario~\citep{Pandolfi:2010dz}. In all our previous results, we have restricted ourselves to parameterize the reionization history in terms of the optical depth to reionization, see \autoref{eq:reio}. To study the impact of a more general reionization scenario on the tension on $n_{\rm s}$, we have explored, as a  first attempt, the so-called \emph{redshift-symmetric} parameterization, which assumes that the free electron fraction follows a step-like function, taking the recombination leftover value at high redshifts and becoming close to one at low redshifts, and being described by the hyperbolic tangent function~\citep{Lewis:2008wr}
\begin{equation}
x_e^{\rm tanh}(z) = \frac{1+f_{\rm He}}{2} \left(1+ \tanh \left[ \frac{y(z_{\rm{re}})-y(z)}{\Delta y} \right] \right),
\label{eqn:tanh}
\end{equation}
where $f_{\rm He}=n_{\rm{He}}/n_{\rm{H}}$ is the Helium fraction, $y(z)=(1+z)^{3/2}$, $\Delta y=3/2(1+z_{\rm{re}})^{1/2}\Delta z$, and $\Delta z$ is the width of the transition. Therefore, the free parameters in this simple approach are the reionization redshift $z_{\rm{re}}$ and  $\Delta z$. However, this reionization scenario renders the very same results, and the $\sim 3\sigma$ tension on $n_{\rm s}$ still persists both with and without the inclusion of large scale structure data, see the right panel of~\autoref{fig:fig3}. More general reionization schemes, such as a Principal Component Analysis (PCA) approach of Refs.~\citep{Hu:2003gh,Mortonson:2007hq,Mortonson:2007tb,Mortonson:2008rx,Mortonson:2009qv,Mortonson:2009xk,Mitra:2010sr,Villanueva-Domingo:2017ahx}  or non-parametric forms for the free electron fraction $x_e(z)$, which is instead described using the function values $x_e(z_i)$ in a number $n$ of fixed redshift points $z_1,\ \ldots,\ z_n$,  are promising and viable phenomenological alternatives that will be further explored in future work.

\subsection{Extending the cosmological model}

\begin{figure}
\includegraphics[width= \columnwidth]{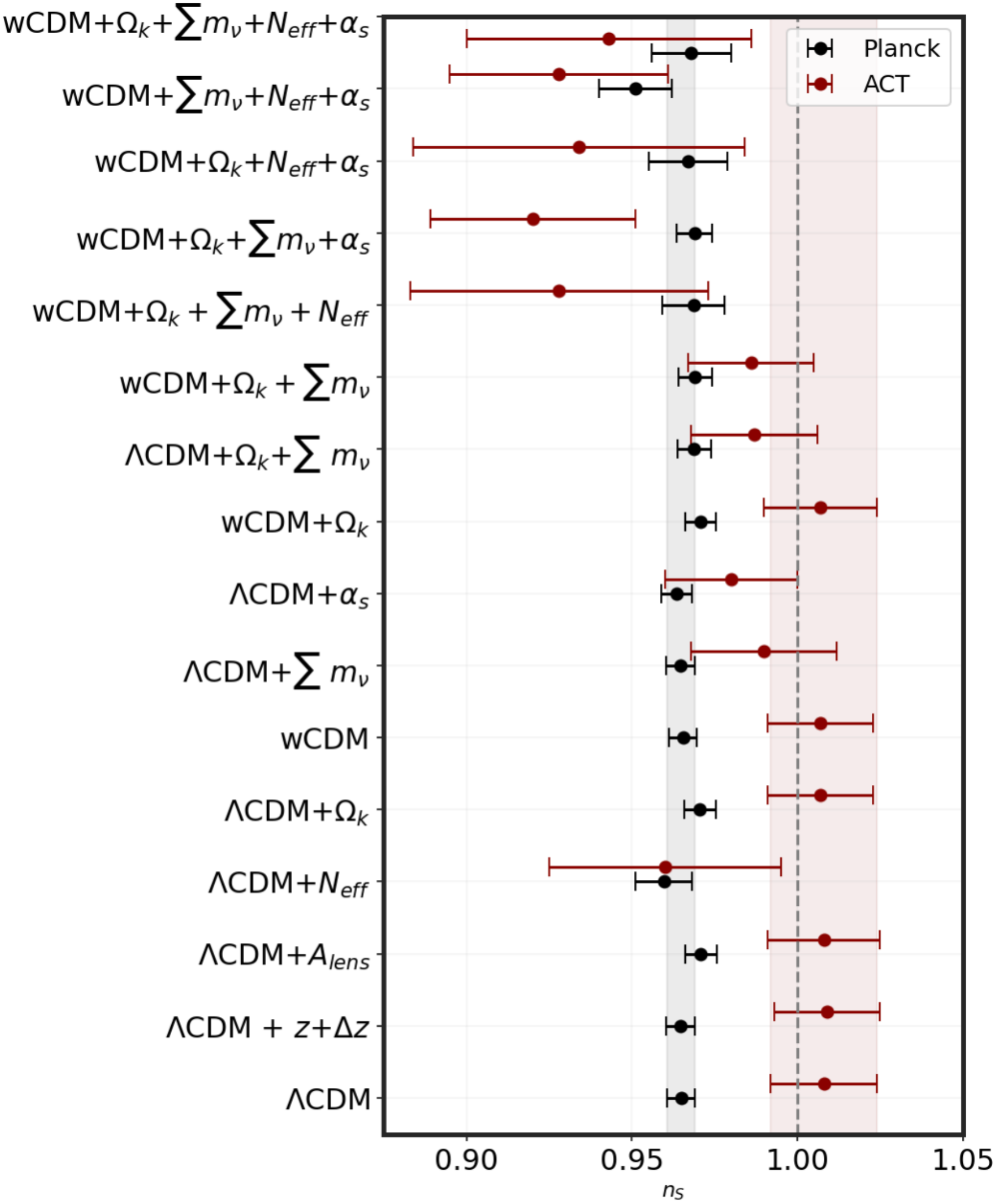}
\caption{\small Mean values and 1$\sigma$ errors  for the scalar spectral index for various extensions of the fiducial cosmology as inferred by \emph{Planck} (black points) and ACT (red points).}
\label{fig:fig4}
\end{figure}

Yet another possibility usually explored when finding anomalies in the cosmological parameter values when combining different data sets is to extend the minimal cosmological model. We have performed many tests in this directions that are summarized in \autoref{fig:fig4} and \autoref{tab.Results.Models}. 

Relaxing the dark energy sector physics alleviates some discrepancies, such as the Hubble constant one  (see Refs.~\citep{Verde:2019ivm,Knox:2019rjx,DiValentino:2020zio,DiValentino:2021izs,Abdalla:2022yfr} and references therein). We have therefore allowed the dark energy equation of state $w$ to be a free parameter, finding  values of $n_\mathrm{s} = 1.010\pm 0.014$ ($n_{\rm s} = 0.9654\pm 0.0042$) for ACT with a prior of $\tau_{\rm reio}=0.065 \pm 0.015$ + BAO DR12 (\emph{Planck} + BAO DR12). The tension is enhanced to the $3.2\sigma$ level. 

Relaxing the assumption of flatness provides a solution to the aforementioned \emph{Planck} lensing anomaly~\citep{DiValentino:2019qzk}. Motivated by this neat result, we have explored here the possibility of having a non-zero curvature parameter $\Omega_k$.  While the \emph{Planck} (TT TE EE) data show a definite preference for a closed Universe at more than $99\%$~CL~\citep{DiValentino:2022oon}, as already mentioned in the introduction, ACT is perfectly consistent with the inflationary prediction for a flat Universe, i.e. with $\Omega_k=0$. More concretely, we find $\Omega_k=-0.0011^{+0.014}_{-0.0093}$ for the case of ACT plus a prior on $\tau_{\rm reio}=0.0544\pm 0.0070$~\citep{Planck:2018vyg}. A very similar conclusion is achieved if the reionization prior considered is $\tau_{\rm reio}=0.065 \pm 0.015$~\citep{ACT:2020gnv} or low multipole polarization data from \emph{Planck} is used. Last, but not least, ACT data is perfectly compatible with $A_\mathrm{lens}=1$~\citep{DiValentino:2022rdg}, as $A_\mathrm{lens}= 0.984^{+0.082}_{-0.094}$ for the combination of ACT plus a prior $\tau_{\rm reio}=0.0544\pm 0.0070$~\citep{Planck:2018vyg}. 

However, none of these extensions have been successful in resolving the discrepancy between the scalar spectral index as measured by \emph{Planck} and ACT, see also \autoref{tab.Results.Models} and \autoref{fig:fig4}.Conversely, as demonstrated in \autoref{fig:fig4}, including the effective number of relativistic degrees of freedom $N_{\rm eff}$ in the sample is the one of the few minimal extensions able to improve the agreement between the two experiments. The reduction in the global tension between ACT and \emph{Planck} has been extensively explored in recent studies~\citep{DiValentino:2022rdg} and found to be closely related to the ACT preference for a value of the effective number of relativistic particles ($N_{\rm eff}=2.35^{+0.40}_{-0.47}$ at 68\%~CL) significantly lower than the value expected within the standard model of particle physics ($N_{\rm eff}=3.044$). This appears to be an important factor also in restoring the concordance for the inflationary predictions since leaving $N_{\rm eff}$ a free parameter in the sample, from ACT we obtain $n_s=0.960\pm0.035$ that is now in perfect agreement with \emph{Planck}. Despite the loss of constraining power on $n_s$ due to the geometrical degeneracy between these two parameters, the agreement is restored because of an actual shift in the mean value of $n_s$ rather than to larger error-bars, see also \autoref{fig:fig4}.

A second interesting avenue to reconcile the disagreement in the value of the spectral index may also rely in modifications to the neutrino sector.
Specifically, adding the total neutrino mass in the sample, from ACT we observe a shift in the value of $n_s$ towards the \emph{Planck} result, and the tension barely reaches the statistical level of 1$\sigma$. However, also in this case, the $n_s$-problem seems to be linked to the anomalous ACT preference for larger values of the neutrino mass ($\sum m_{\nu} \lesssim 1$ eV), generally at odd with the Planck and BAO cosmological measurements, that are instead disfavoring the inverted ordering as the one governing the mass pattern of neutral fermions~\citep{DiValentino:2021hoh}. Indeed, due to the strong anti-correlation between these two parameters, larger neutrino masses allow to recover values of $n_s$ closer to those predicted by Planck.

\begin{figure}
\includegraphics[width= \columnwidth]{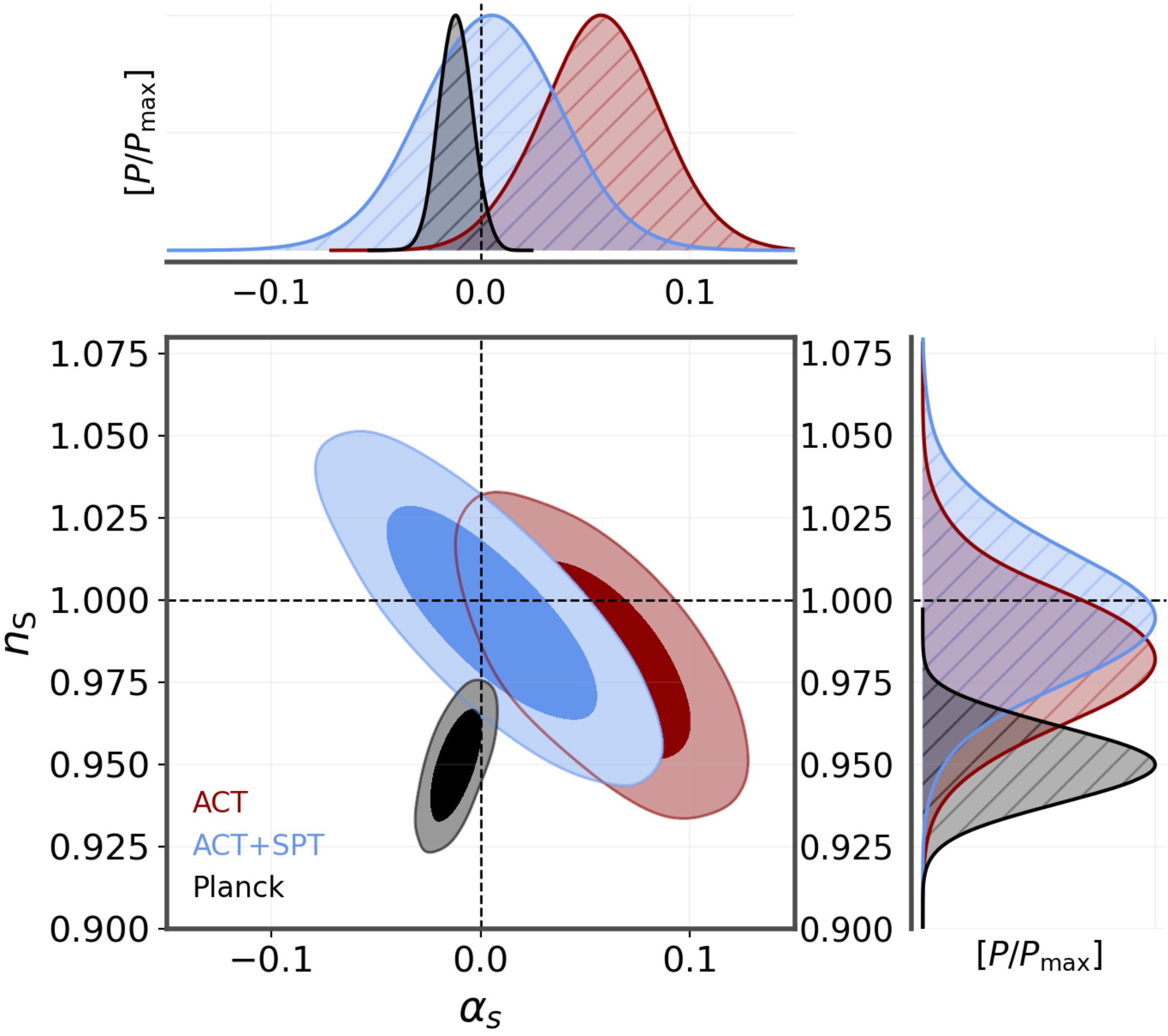}
\caption{\small One-dimensional posterior distributions and two-dimensional joint marginalized contours for the spectral index $n_s$ and its running $\alpha_s=dn_{s}/d\log k$ inferred by the \emph{Planck} satellite and the Atacama Cosmology Telescope, both alone and in combination with the South Pole Telescope polarization measurements.}
\label{fig:fig5}
\end{figure}

Another important clue that can be inferred from \autoref{fig:fig4} is the tendency for the value of $n_s$ predicted by ACT to decrease significantly when extending the inflationary scenario by accounting also for the possibility of a running in the scalar spectral index $\alpha_{\rm s}= d n_{\rm s} / d\log k$. \autoref{fig:fig5} depicts the impact of considering a running of the scalar spectral index in the parameter allowed regions for the cases of \emph{Planck} and ACT data, either alone or combined with SPT polarization measurements. In the case of CMB data alone, we obtain $n_{\rm s} = 0.950\pm 0.011$ ($n_\mathrm{s} = 0.982\pm 0.020$) for \emph{Planck} (ACT) data, leading to a very mild $1.6\sigma$ discrepancy. If the BAO (DR12 or DR16) dataset is also considered, the disagreement gets further diluted and it barely reaches the $1\sigma$ significance. However, the tension in $n_{\rm s}$ maps into a controversy in the values of $\alpha_{\rm s}$: while \emph{Planck} alone prefers a negative value of $\alpha_{\rm s}= -0.0119\pm 0.0079$, ACT measurements favor a positive running $\alpha_{\rm s}= 0.058\pm 0.028$, leading to a $2.5\sigma$ tension. The addition of BAO does not modify this result: a positive running can contribute to positively tilt the spectrum and mimic the effect of a larger $n_s$, see also the strong positive correlation between these two parameters in \autoref{fig:fig5}. The preference for a positive tilted spectrum from small scale CMB observations, which is shown to persist even combining ACT with WMAP 9-year observations~\citep{Forconi:2021que}, challenges canonical inflationary scenarios, as the predictions from all these models provide a negative value of $\alpha_{\rm s}$, see e.g.~\cite{Martin:2013tda,Escudero:2015wba,Planck:2018jri}. To further understand this discrepancy, we reconsider to neglect the ACT polarization data by replacing it with SPT polarization TE and EE measurements, as done for the baseline $\Lambda$CDM scenario. Combining these two small-scale CMB observations, the spectral index remains centered around $n_s\sim 1$ but the decrease in constraining power weakens the tension with \emph{Planck}. However, it is noteworthy that disregarding the ACT polarization data, the value of $\alpha_{\rm s}$ shifts towards $\alpha_s\simeq 0$, lending support to the hypothesis that the ACT polarization measurements may play a prominent role in producing this unexpected result.

Finally, as seen in \autoref{fig:fig4}, the disparities in $n_s$ experience a noticeable decline as more parameters are added to the background cosmology. This improvement can be partially attributed to the decreased data constraining power that commonly results from incorporating a high number of parameters, leading to geometric degeneracies among them. However, it can be noticed that ACT data exhibits a trend towards smaller values of $n_s$, which is not evident in the \emph{Planck} data. This deviation is primarily driven by the combined effect of other anomalous parameters discussed so far, most notably $N_{\rm eff}$, $\sum m_{\nu}$ and $\alpha_s$. Hence, while it is entirely possible that the discrepancy in the spectral index may reflect important observational systematic errors, we cannot rule out that this difference may stem from a limitation in the standard cosmological model to accurately reflect small scale (high multipoles) CMB observations as probed by ACT.

\section{Conclusion}
\label{sec.5}

Both the \emph{Planck} satellite and the Atacama Cosmology Telescope show intriguing anomalies that seem to challenge the typical predictions of inflationary theories: the first data-set seems to disfavor the inflationary prediction for a flat background geometry at more than 99.9 \% CL while the second, albeit in perfect agreement with spatial flatness, shows a preference for a larger spectral index consistent with a Harrison-Zel’dovich scale-invariant spectrum ($n_s=1$) of primordial density perturbations, introducing a tension with a significance of 99.3\% CL with the results from the \emph{Planck} satellite. These anomalies suggest either the presence of important observational systematic errors in one or both data-sets or a departure from the theoretical framework. In this work we have extensively explored both possibilities, extending the analysis presented in Ref.~\cite{ACT:2020gnv}.

Concerning the possibility of observational systematics in the observations, we have performed several tests in this direction. First and foremost, our analysis definitively proves that this preference remains robust with the addition of large scale structure information data both in the form of BAO (DR12 and DR16) measurements and shear-shear, galaxy-galaxy, and galaxy-shear correlation functions from the first year of the Dark Energy Survey, see \autoref{fig:fig2}. In addition we have demonstrated that the inclusion of low multipole polarization data from the \emph{Planck} measurements of E-modes at multipoles $2\le \ell \le 30$, while breaking the degeneracy between the optical depth at reionization ($\tau$) and the scalar spectral index, is not able to explain this anomaly. Finally, following Ref.~\cite{ACT:2020gnv} where it was argued that an overall TE calibration could eventually explain the mismatch in $n_s$, we have neglected any information arising from ACT polarization measurements (TE EE) and combined ACT temperature anisotropies (TT) with SPT polarization data (TE EE). Although the combination of ACT and SPT appears to restore the agreement for $\Omega_b\,h^2$ and reduce the disagreement on $n_s$ at the level of two standard deviations (suggesting the possibility of systematic in the ACT data), this result is mostly due to the loss of constraining power rather than an actual shift in the value of $n_s$, see also \autoref{fig:fig1}. Therefore restoring the agreement for $\Omega_b\,h^2$ doesn't seems enough to reconcile the $n_s$ discrepancy.

Given that none of the tests has been conclusive in explaining this unexpected result,  we have investigated several possible theoretical explanations behind the disagreement. As a first attempt, we have relaxed the canonical reionization scenario and explored the so-called \emph{redshift-symmetric} parameterization, which assumes that the free electron fraction follows a step-like function, taking the recombination leftover value at high redshifts. While it is not excluded that the disagreement may actually be reduced in alternative reionization scenarios, this particular parametrization has proved to be largely unable to explain the tension on $n_s$, see \autoref{fig:fig3}. Another possibility usually explored when finding anomalies in the cosmological parameter values is to extend the minimal cosmological model. We have therefore explored over a dozen of extended cosmologies, summarized in \autoref{fig:fig4}. We have argued that the $n_s$-tensions appears to be strongly related to the anomalous behavior of other parameters typically fixed at $\Lambda$CDM but that, when fitted with ACT data, in turn show significant deviations from the baseline cosmology. Remarkably, modifications to the neutrino sectors, parameterized both in terms of the effective number of relativistic neutrinos $N_{\rm eff}$ and the total neutrino mass $\sum m_{\nu}$, are able to significantly reduce the disagreement on $n_s$. However, they resulting in a preference of the ACT data for a value $N_{\rm eff}$ significantly lower than expected within the standard model of particle physics, and in a large neutrino mass at odds with global fit analyses~\cite{deSalas:2020pgw}. Yet another possible phenomenological avenue to settle this issue includes a possible positive running of the scalar spectral index $\alpha_{\rm s}>0$. In this latter case the tension is translated into a discrepancy on $\alpha_{\rm s}$ with \emph{Planck} preferring a negative running and ACT a positive one. It is also noteworthy that the combination ACT+SPT does not lead to any evidence for a running spectral index, see \autoref{fig:fig5}. From the theoretical/model-building approach, non-standard inflation theories may also provide a solution, while being testable by near future CMB B-mode experiments.

In conclusion, while our analysis of the ACT and SPT temperature and polarization data suggests that the disagreement in the spectral index may be due to systematic errors in the observations, we cannot disregard the possibility that the tension is rooted in the limitations of the standard cosmological model. Our findings indicate that the value of the spectral index measured by ACT is highly dependent on the underlying assumptions of the $\Lambda$CDM cosmology. This raises the possibility that the standard cosmological model is not fully modeled to accurately represent the small scale (high multipole) observations of the CMB as probed by ACT, which could account for the discrepancy in the spectral index.

\section*{Acknowledgements}
This work has been partially supported by the MCIN/AEI/10.13039/501100011033 of Spain under grant PID2020-113644GB-I00, by the Generalitat Valenciana of Spain under grants PROMETEO/2019/083, PROMETEO/2021/087 and by the European Union’s Framework Programme for Research and Innovation Horizon 2020 (2014–2020) under grant H2020-MSCA-ITN-2019/860881-HIDDeN.
EDV is supported by a Royal Society Dorothy Hodgkin Research Fellowship. FR acknowledges support from the NWO and the Dutch Ministry of Education, Culture and Science (OCW) (through NWO VIDI Grant No.2019/ENW/00678104 and from the D-ITP consortium) WG and AM are supported by "Theoretical Astroparticle Physics" (TAsP), iniziativa specifica INFN.
    	
\section*{Data availability}
The data underlying this article will be shared on reasonable request to the corresponding author.

\bibliographystyle{mnras}
\bibliography{biblio} 

\begin{thebibliography}{}
\makeatletter
\relax
\def\mn@urlcharsother{\let\do\@makeother \do\$\do\&\do\#\do\^\do\_\do\%\do\~}
\def\mn@doi{\begingroup\mn@urlcharsother \@ifnextchar [ {\mn@doi@}
  {\mn@doi@[]}}
\def\mn@doi@[#1]#2{\def\@tempa{#1}\ifx\@tempa\@empty \href
  {http://dx.doi.org/#2} {doi:#2}\else \href {http://dx.doi.org/#2} {#1}\fi
  \endgroup}
\def\mn@eprint#1#2{\mn@eprint@#1:#2::\@nil}
\def\mn@eprint@arXiv#1{\href {http://arxiv.org/abs/#1} {{\tt arXiv:#1}}}
\def\mn@eprint@dblp#1{\href {http://dblp.uni-trier.de/rec/bibtex/#1.xml}
  {dblp:#1}}
\def\mn@eprint@#1:#2:#3:#4\@nil{\def\@tempa {#1}\def\@tempb {#2}\def\@tempc
  {#3}\ifx \@tempc \@empty \let \@tempc \@tempb \let \@tempb \@tempa \fi \ifx
  \@tempb \@empty \def\@tempb {arXiv}\fi \@ifundefined
  {mn@eprint@\@tempb}{\@tempb:\@tempc}{\expandafter \expandafter \csname
  mn@eprint@\@tempb\endcsname \expandafter{\@tempc}}}

\bibitem[\protect\citeauthoryear{Abbott et~al.}{Abbott
  et~al.}{2018}]{DES:2017myr}
Abbott T. M.~C.,  et~al., 2018, \mn@doi [Phys. Rev. D]
  {10.1103/PhysRevD.98.043526}, 98, 043526

\bibitem[\protect\citeauthoryear{Abbott et~al.}{Abbott
  et~al.}{2022}]{DES:2021wwk}
Abbott T. M.~C.,  et~al., 2022, \mn@doi [Phys. Rev. D]
  {10.1103/PhysRevD.105.023520}, 105, 023520

\bibitem[\protect\citeauthoryear{Abdalla et~al.}{Abdalla
  et~al.}{2022}]{Abdalla:2022yfr}
Abdalla E.,  et~al., 2022, \mn@doi [JHEAp] {10.1016/j.jheap.2022.04.002}, 34,
  49

\bibitem[\protect\citeauthoryear{Aghanim et~al.}{Aghanim
  et~al.}{2020a}]{Planck:2018nkj}
Aghanim N.,  et~al., 2020a, \mn@doi [Astron. Astrophys.]
  {10.1051/0004-6361/201833880}, 641, A1

\bibitem[\protect\citeauthoryear{Aghanim et~al.}{Aghanim
  et~al.}{2020b}]{Planck:2019nip}
Aghanim N.,  et~al., 2020b, \mn@doi [Astron. Astrophys.]
  {10.1051/0004-6361/201936386}, 641, A5

\bibitem[\protect\citeauthoryear{Aghanim et~al.}{Aghanim
  et~al.}{2020c}]{Planck:2018vyg}
Aghanim N.,  et~al., 2020c, \mn@doi [Astron. Astrophys.]
  {10.1051/0004-6361/201833910}, 641, A6

\bibitem[\protect\citeauthoryear{Aiola et~al.}{Aiola
  et~al.}{2020}]{ACT:2020gnv}
Aiola S.,  et~al., 2020, \mn@doi [JCAP] {10.1088/1475-7516/2020/12/047}, 12,
  047

\bibitem[\protect\citeauthoryear{Akrami et~al.}{Akrami
  et~al.}{2020}]{Planck:2018jri}
Akrami Y.,  et~al., 2020, \mn@doi [Astron. Astrophys.]
  {10.1051/0004-6361/201833887}, 641, A10

\bibitem[\protect\citeauthoryear{Barrow}{Barrow}{1990}]{Barrow:1990vx}
Barrow J.~D.,  1990, \mn@doi [Phys. Lett. B] {10.1016/0370-2693(90)90093-L},
  235, 40

\bibitem[\protect\citeauthoryear{Barrow \& Liddle}{Barrow \&
  Liddle}{1993}]{Barrow:1993zq}
Barrow J.~D.,  Liddle A.~R.,  1993, \mn@doi [Phys. Rev. D]
  {10.1103/PhysRevD.47.R5219}, 47, R5219

\bibitem[\protect\citeauthoryear{Barrow \& Saich}{Barrow \&
  Saich}{1990}]{Barrow:1990td}
Barrow J.~D.,  Saich P.,  1990, \mn@doi [Phys. Lett. B]
  {10.1016/0370-2693(90)91007-X}, 249, 406

\bibitem[\protect\citeauthoryear{Barrow, Liddle  \& Pahud}{Barrow
  et~al.}{2006}]{Barrow:2006dh}
Barrow J.~D.,  Liddle A.~R.,   Pahud C.,  2006, \mn@doi [Phys. Rev. D]
  {10.1103/PhysRevD.74.127305}, 74, 127305

\bibitem[\protect\citeauthoryear{Calder\'on, Shafieloo, Hazra  \&
  Sohn}{Calder\'on et~al.}{2023}]{Calderon:2023obf}
Calder\'on R.,  Shafieloo A.,  Hazra D.~K.,   Sohn W.,  2023, {On the
  consistency of $\Lambda$CDM with CMB measurements in light of the latest
  Planck, ACT, and SPT data} (\mn@eprint {arXiv} {2302.14300})

\bibitem[\protect\citeauthoryear{Choi et~al.}{Choi et~al.}{2020}]{ACT:2020frw}
Choi S.~K.,  et~al., 2020, \mn@doi [JCAP] {10.1088/1475-7516/2020/12/045}, 12,
  045

\bibitem[\protect\citeauthoryear{Dawson et~al.}{Dawson
  et~al.}{2013}]{BOSS:2012dmf}
Dawson K.~S.,  et~al., 2013, \mn@doi [Astron. J.] {10.1088/0004-6256/145/1/10},
  145, 10

\bibitem[\protect\citeauthoryear{Dawson et~al.}{Dawson
  et~al.}{2016}]{Dawson:2015wdb}
Dawson K.~S.,  et~al., 2016, \mn@doi [Astron. J.] {10.3847/0004-6256/151/2/44},
  151, 44

\bibitem[\protect\citeauthoryear{Di~Valentino, Melchiorri, Fantaye  \&
  Heavens}{Di~Valentino et~al.}{2018}]{DiValentino:2018zjj}
Di~Valentino E.,  Melchiorri A.,  Fantaye Y.,   Heavens A.,  2018, \mn@doi
  [Phys. Rev. D] {10.1103/PhysRevD.98.063508}, 98, 063508

\bibitem[\protect\citeauthoryear{Di~Valentino, Melchiorri  \&
  Silk}{Di~Valentino et~al.}{2019}]{DiValentino:2019qzk}
Di~Valentino E.,  Melchiorri A.,   Silk J.,  2019, \mn@doi [Nature Astron.]
  {10.1038/s41550-019-0906-9}, 4, 196

\bibitem[\protect\citeauthoryear{Di~Valentino et~al.,}{Di~Valentino
  et~al.}{2021a}]{DiValentino:2021izs}
Di~Valentino E.,  et~al., 2021a, \mn@doi [Class. Quant. Grav.]
  {10.1088/1361-6382/ac086d}, 38, 153001

\bibitem[\protect\citeauthoryear{Di~Valentino, Gariazzo  \& Mena}{Di~Valentino
  et~al.}{2021b}]{DiValentino:2021hoh}
Di~Valentino E.,  Gariazzo S.,   Mena O.,  2021b, \mn@doi [Phys. Rev. D]
  {10.1103/PhysRevD.104.083504}, 104, 083504

\bibitem[\protect\citeauthoryear{Di~Valentino et~al.}{Di~Valentino
  et~al.}{2021c}]{DiValentino:2020vvd}
Di~Valentino E.,  et~al., 2021c, \mn@doi [Astropart. Phys.]
  {10.1016/j.astropartphys.2021.102604}, 131, 102604

\bibitem[\protect\citeauthoryear{Di~Valentino et~al.}{Di~Valentino
  et~al.}{2021d}]{DiValentino:2020zio}
Di~Valentino E.,  et~al., 2021d, \mn@doi [Astropart. Phys.]
  {10.1016/j.astropartphys.2021.102605}, 131, 102605

\bibitem[\protect\citeauthoryear{Di~Valentino, Melchiorri  \&
  Silk}{Di~Valentino et~al.}{2021e}]{DiValentino:2020hov}
Di~Valentino E.,  Melchiorri A.,   Silk J.,  2021e, \mn@doi [Astrophys. J.
  Lett.] {10.3847/2041-8213/abe1c4}, 908, L9

\bibitem[\protect\citeauthoryear{Di~Valentino, Giar\`e, Melchiorri  \&
  Silk}{Di~Valentino et~al.}{2022}]{DiValentino:2022oon}
Di~Valentino E.,  Giar\`e W.,  Melchiorri A.,   Silk J.,  2022, \mn@doi [Phys.
  Rev. D] {10.1103/PhysRevD.106.103506}, 106, 103506

\bibitem[\protect\citeauthoryear{Di~Valentino, Giar\`e, Melchiorri  \&
  Silk}{Di~Valentino et~al.}{2023}]{DiValentino:2022rdg}
Di~Valentino E.,  Giar\`e W.,  Melchiorri A.,   Silk J.,  2023, \mn@doi [Mon.
  Not. Roy. Astron. Soc.] {10.1093/mnras/stad152}, 520, 210

\bibitem[\protect\citeauthoryear{Escudero, Ram\'\i{}rez, Boubekeur, Giusarma
  \& Mena}{Escudero et~al.}{2016}]{Escudero:2015wba}
Escudero M.,  Ram\'\i{}rez H.,  Boubekeur L.,  Giusarma E.,   Mena O.,  2016,
  \mn@doi [JCAP] {10.1088/1475-7516/2016/02/020}, 02, 020

\bibitem[\protect\citeauthoryear{Forconi, Giar\`e, Di~Valentino  \&
  Melchiorri}{Forconi et~al.}{2021}]{Forconi:2021que}
Forconi M.,  Giar\`e W.,  Di~Valentino E.,   Melchiorri A.,  2021, \mn@doi
  [Phys. Rev. D] {10.1103/PhysRevD.104.103528}, 104, 103528

\bibitem[\protect\citeauthoryear{Handley}{Handley}{2021}]{Handley:2019tkm}
Handley W.,  2021, \mn@doi [Phys. Rev. D] {10.1103/PhysRevD.103.L041301}, 103,
  L041301

\bibitem[\protect\citeauthoryear{Harrison}{Harrison}{1970}]{Harrison:1969fb}
Harrison E.~R.,  1970, \mn@doi [Phys. Rev. D] {10.1103/PhysRevD.1.2726}, 1,
  2726

\bibitem[\protect\citeauthoryear{Heymans et~al.}{Heymans
  et~al.}{2021}]{Heymans:2020gsg}
Heymans C.,  et~al., 2021, \mn@doi [Astron. Astrophys.]
  {10.1051/0004-6361/202039063}, 646, A140

\bibitem[\protect\citeauthoryear{Hinshaw et~al.}{Hinshaw
  et~al.}{2013}]{WMAP:2012nax}
Hinshaw G.,  et~al., 2013, \mn@doi [Astrophys. J. Suppl.]
  {10.1088/0067-0049/208/2/19}, 208, 19

\bibitem[\protect\citeauthoryear{Howlett, Lewis, Hall  \& Challinor}{Howlett
  et~al.}{2012}]{Howlett:2012mh}
Howlett C.,  Lewis A.,  Hall A.,   Challinor A.,  2012, \mn@doi [JCAP]
  {10.1088/1475-7516/2012/04/027}, 04, 027

\bibitem[\protect\citeauthoryear{Hu \& Holder}{Hu \& Holder}{2003}]{Hu:2003gh}
Hu W.,  Holder G.~P.,  2003, \mn@doi [Phys. Rev. D]
  {10.1103/PhysRevD.68.023001}, 68, 023001

\bibitem[\protect\citeauthoryear{Jiang \& Piao}{Jiang \&
  Piao}{2022}]{Jiang:2022uyg}
Jiang J.-Q.,  Piao Y.-S.,  2022, \mn@doi [Phys. Rev. D]
  {10.1103/PhysRevD.105.103514}, 105, 103514

\bibitem[\protect\citeauthoryear{Jiang, Ye  \& Piao}{Jiang
  et~al.}{2022}]{Jiang:2022qlj}
Jiang J.-Q.,  Ye G.,   Piao Y.-S.,  2022, {Return of Harrison-Zeldovich
  spectrum in light of recent cosmological tensions} (\mn@eprint {arXiv}
  {2210.06125})

\bibitem[\protect\citeauthoryear{Knox \& Millea}{Knox \&
  Millea}{2020}]{Knox:2019rjx}
Knox L.,  Millea M.,  2020, \mn@doi [Phys. Rev. D]
  {10.1103/PhysRevD.101.043533}, 101, 043533

\bibitem[\protect\citeauthoryear{Lewis}{Lewis}{2008}]{Lewis:2008wr}
Lewis A.,  2008, \mn@doi [Phys. Rev. D] {10.1103/PhysRevD.78.023002}, 78,
  023002

\bibitem[\protect\citeauthoryear{Lewis \& Bridle}{Lewis \&
  Bridle}{2002}]{Lewis:2002ah}
Lewis A.,  Bridle S.,  2002, \mn@doi [Phys. Rev. D]
  {10.1103/PhysRevD.66.103511}, 66, 103511

\bibitem[\protect\citeauthoryear{Lewis, Challinor  \& Lasenby}{Lewis
  et~al.}{2000}]{Lewis:1999bs}
Lewis A.,  Challinor A.,   Lasenby A.,  2000, \mn@doi [Astrophys. J.]
  {10.1086/309179}, 538, 473

\bibitem[\protect\citeauthoryear{Lin}{Lin}{2022}]{Lin:2022gbl}
Lin C.-M.,  2022, {Revisiting D-term inflation with gravitational waves and
  Hubble tension} (\mn@eprint {arXiv} {2204.10475})

\bibitem[\protect\citeauthoryear{Martin, Ringeval  \& Vennin}{Martin
  et~al.}{2014}]{Martin:2013tda}
Martin J.,  Ringeval C.,   Vennin V.,  2014, \mn@doi [Phys. Dark Univ.]
  {10.1016/j.dark.2014.01.003}, 5-6, 75

\bibitem[\protect\citeauthoryear{Mitra, Choudhury  \& Ferrara}{Mitra
  et~al.}{2011}]{Mitra:2010sr}
Mitra S.,  Choudhury T.~R.,   Ferrara A.,  2011, \mn@doi [Mon. Not. Roy.
  Astron. Soc.] {10.1111/j.1365-2966.2011.18234.x}, 413, 1569

\bibitem[\protect\citeauthoryear{Mortonson \& Hu}{Mortonson \&
  Hu}{2008a}]{Mortonson:2007tb}
Mortonson M.~J.,  Hu W.,  2008a, \mn@doi [Phys. Rev. D]
  {10.1103/PhysRevD.77.043506}, 77, 043506

\bibitem[\protect\citeauthoryear{Mortonson \& Hu}{Mortonson \&
  Hu}{2008b}]{Mortonson:2007hq}
Mortonson M.~J.,  Hu W.,  2008b, \mn@doi [Astrophys. J.] {10.1086/523958}, 672,
  737

\bibitem[\protect\citeauthoryear{Mortonson \& Hu}{Mortonson \&
  Hu}{2008c}]{Mortonson:2008rx}
Mortonson M.~J.,  Hu W.,  2008c, \mn@doi [Astrophys. J. Lett.]
  {10.1086/593031}, 686, L53

\bibitem[\protect\citeauthoryear{Mortonson \& Hu}{Mortonson \&
  Hu}{2009}]{Mortonson:2009xk}
Mortonson M.~J.,  Hu W.,  2009, \mn@doi [Phys. Rev. D]
  {10.1103/PhysRevD.80.027301}, 80, 027301

\bibitem[\protect\citeauthoryear{Mortonson, Dvorkin, Peiris  \& Hu}{Mortonson
  et~al.}{2009}]{Mortonson:2009qv}
Mortonson M.~J.,  Dvorkin C.,  Peiris H.~V.,   Hu W.,  2009, \mn@doi [Phys.
  Rev. D] {10.1103/PhysRevD.79.103519}, 79, 103519

\bibitem[\protect\citeauthoryear{Motloch \& Hu}{Motloch \&
  Hu}{2018}]{Motloch:2018pjy}
Motloch P.,  Hu W.,  2018, \mn@doi [Phys. Rev. D] {10.1103/PhysRevD.97.103536},
  97, 103536

\bibitem[\protect\citeauthoryear{{Neal}}{{Neal}}{2005}]{Neal:2005}
{Neal} R.~M.,  2005, ArXiv Mathematics e-prints, \href
  {http://adsabs.harvard.edu/abs/2005math......2099N} {}

\bibitem[\protect\citeauthoryear{Pandolfi, Cooray, Giusarma, Kolb, Melchiorri,
  Mena  \& Serra}{Pandolfi et~al.}{2010}]{Pandolfi:2010dz}
Pandolfi S.,  Cooray A.,  Giusarma E.,  Kolb E.~W.,  Melchiorri A.,  Mena O.,
  Serra P.,  2010, \mn@doi [Phys. Rev. D] {10.1103/PhysRevD.81.123509}, 81,
  123509

\bibitem[\protect\citeauthoryear{Peebles \& Yu}{Peebles \&
  Yu}{1970}]{Peebles:1970ag}
Peebles P. J.~E.,  Yu J.~T.,  1970, \mn@doi [Astrophys. J.] {10.1086/150713},
  162, 815

\bibitem[\protect\citeauthoryear{Perivolaropoulos \& Skara}{Perivolaropoulos \&
  Skara}{2022}]{Perivolaropoulos:2021jda}
Perivolaropoulos L.,  Skara F.,  2022, \mn@doi [New Astron. Rev.]
  {10.1016/j.newar.2022.101659}, 95, 101659

\bibitem[\protect\citeauthoryear{Riess et~al.}{Riess
  et~al.}{2022}]{Riess:2021jrx}
Riess A.~G.,  et~al., 2022, \mn@doi [Astrophys. J. Lett.]
  {10.3847/2041-8213/ac5c5b}, 934, L7

\bibitem[\protect\citeauthoryear{Semenaite et~al.,}{Semenaite
  et~al.}{2022}]{Semenaite:2022unt}
Semenaite A.,  et~al., 2022, {Beyond $\Lambda$CDM constraints from the full
  shape clustering measurements from BOSS and eBOSS} (\mn@eprint {arXiv}
  {2210.07304})

\bibitem[\protect\citeauthoryear{Starobinsky}{Starobinsky}{2005}]{Starobinsky:2005ab}
Starobinsky A.~A.,  2005, \mn@doi [JETP Lett.] {10.1134/1.2121807}, 82, 169

\bibitem[\protect\citeauthoryear{Takahashi \& Yin}{Takahashi \&
  Yin}{2022}]{Takahashi:2021bti}
Takahashi F.,  Yin W.,  2022, \mn@doi [Phys. Lett. B]
  {10.1016/j.physletb.2022.137143}, 830, 137143

\bibitem[\protect\citeauthoryear{Torrado \& Lewis}{Torrado \&
  Lewis}{2020}]{Torrado:2020xyz}
Torrado J.,  Lewis A.,  2020, arXiv:2005.05290

\bibitem[\protect\citeauthoryear{Vallinotto, Copeland, Kolb, Liddle  \&
  Steer}{Vallinotto et~al.}{2004}]{Vallinotto:2003vf}
Vallinotto A.,  Copeland E.~J.,  Kolb E.~W.,  Liddle A.~R.,   Steer D.~A.,
  2004, \mn@doi [Phys. Rev. D] {10.1103/PhysRevD.69.103519}, 69, 103519

\bibitem[\protect\citeauthoryear{Verde, Treu  \& Riess}{Verde
  et~al.}{2019}]{Verde:2019ivm}
Verde L.,  Treu T.,   Riess A.~G.,  2019, \mn@doi [Nature Astron.]
  {10.1038/s41550-019-0902-0}, 3, 891

\bibitem[\protect\citeauthoryear{Villanueva-Domingo, Gariazzo, Gnedin  \&
  Mena}{Villanueva-Domingo et~al.}{2018}]{Villanueva-Domingo:2017ahx}
Villanueva-Domingo P.,  Gariazzo S.,  Gnedin N.~Y.,   Mena O.,  2018, \mn@doi
  [JCAP] {10.1088/1475-7516/2018/04/024}, 04, 024

\bibitem[\protect\citeauthoryear{Ye, Jiang  \& Piao}{Ye
  et~al.}{2022}]{Ye:2022efx}
Ye G.,  Jiang J.-Q.,   Piao Y.-S.,  2022, {Towards hybrid inflation with
  $n_s=1$ in light of Hubble tension and primordial gravitational waves}
  (\mn@eprint {arXiv} {2205.02478})

\bibitem[\protect\citeauthoryear{Zeldovich}{Zeldovich}{1972}]{Zeldovich:1972zz}
Zeldovich Y.~B.,  1972, \mn@doi [Mon. Not. Roy. Astron. Soc.]
  {10.1093/mnras/160.1.1P}, 160, 1P

\bibitem[\protect\citeauthoryear{de Salas, Forero, Gariazzo,
  Mart\'\i{}nez-Mirav\'e, Mena, Ternes, T\'ortola  \& Valle}{de~Salas
  et~al.}{2021}]{deSalas:2020pgw}
de Salas P.~F.,  Forero D.~V.,  Gariazzo S.,  Mart\'\i{}nez-Mirav\'e P.,  Mena
  O.,  Ternes C.~A.,  T\'ortola M.,   Valle J. W.~F.,  2021, \mn@doi [JHEP]
  {10.1007/JHEP02(2021)071}, 02, 071

\bibitem[\protect\citeauthoryear{del Campo \& Herrera}{del Campo \&
  Herrera}{2007}]{delCampo:2007iw}
del Campo S.,  Herrera R.,  2007, \mn@doi [Phys. Rev. D]
  {10.1103/PhysRevD.76.103503}, 76, 103503

\makeatother
\end{thebibliography}

\bsp	

\label{lastpage}

\end{document}